\documentclass[pra,aps,twocolumn,nofootinbib,floatfix]{revtex4}
\usepackage{graphicx}
\usepackage{bm}

\def\be{\begin{equation}}
\def\ee{\end{equation}}
\def\ba{\begin{eqnarray}}
\def\ea{\end{eqnarray}}
\def\half{\frac{1}{2}}

\def\th{\theta}
\def\tth{\tilde{\theta}}
\def\eps{\epsilon}
\def\bmi{\bm{i}}
\def\vgs{\vartheta_{gs}}
\def\dtheta{\dot{\theta}}
\def\aneg{\alpha^{(-)}}
\def\e{{\cal E}}
\def\egs{{\cal E}_{gs}}
\def\edrop{{\cal E}_{drop}}
\def\edouble{{\cal E}_{double}}
\def\tS{\widetilde{S}}
\def\Sinst{S_{inst}}
\def\bbif{\beta_{bif}}
\begin{document}
\title{Decay of near-critical currents in superconducting nanowires}
\author{Sergei Khlebnikov}
\email{skhleb@purdue.edu}
\affiliation{Department of Physics and Astronomy, Purdue University, 
West Lafayette, IN 47907, USA}
\begin{abstract}
We consider decay of supercurrent via phase slips in a discrete one-dimensional
superconductor (a chain of nodes connected by superconducting links), aiming to 
explain the experimentally observed power-3/2 
scaling of the activation barrier in nanowires at currents 
close to the critical. We find that, in this discrete model, 
the power-3/2 scaling holds for both long and short wires even
in the presence of bulk superconducting leads, despite the suppression of 
thermal fluctuations at the ends. We also consider
decay via tunneling (quantum phase slips), which becomes important at low temperatures.
We find numerically the relevant Euclidean
solutions (periodic instantons) and determine the scaling of the tunneling exponent 
near the critical current. The scaling law is power-5/4, different from that
of the thermal activation exponent.
\end{abstract}
\maketitle
\setcounter{footnote}{1}
\section{Introduction}
The recent observation in \cite{Belkin&al} of double phase slips  
in superconducting nanowires and the interpretation of these there  
as paired quantum phase slips have brought to the fore the question of to 
what extent a nanowire can
be modeled as a lumped superconducting element, i.e., described by a single phase
variable, similarly to a Josephson junction (JJ). In \cite{Belkin&al}, 
such a description has been 
found to provide good fits to the experimental data. Moreover, in that work, as well as
in some of the earlier experiments \cite{Li&al,Aref&al}, some of the behavior
characteristic
of a JJ has been identified in nanowires also for the classical, thermally activated 
variety of phase slips. This concerns specifically the 
scaling law followed by the activation barrier at currents near the critical.

Recall that, for a JJ with critical current $I_c$, under biasing current 
$I_b$, the activation barrier scales as
\be
\Delta E = \mbox{const} \times (1 - \bmi_b)^{3/2} 
\label{3/2}
\ee
as $\bmi_b = I_b/I_c \to 1$ \cite{Fulton&Dunkleberger}. 
This scaling is different from the one obtained 
for a nanowire (an extended one-dimensional superconductor) in the framework of the
Ginzburg-Landau (GL) theory. In that case, the barrier is
determined by the
energy of the Langer-Ambegaokar (LA) \cite{LA} saddle-point and scales as
$(1 - \bmi_b)^{5/4}$ \cite{Tinkham&al}. Somewhat surprisingly,
as a description of the experimental data
on nanowires, the power-3/2 scaling characteristic of a JJ typically works quite 
well \cite{Belkin&al,Li&al,Aref&al}, although there is one type of wires 
(the crystalline wires
of Ref.~\cite{Aref&al}) where the power-5/4 scaling has been found to work better.

Of course, a priori there is no reason to expect that the GL theory will work at
temperatures where (\ref{3/2}) is typically
observed (which are well below the critical $T_c$),
so one may take the abovementioned experimental results simply as an indication 
that one should use a different model of nanowire.\footnote{
In this respect, it may be significant that the crystalline wires of Ref.
\cite{Aref&al}---the one case where the power-5/4 is found to work better---have 
lower values of $T_c$ than their amorphous counterparts.} 
One alternative
is to consider the wire as a discrete set of nodes connected by
superconducting links, with a phase variable defined on each node.
The distance between the nodes plays the role of the ultraviolet cutoff and
corresponds to the ``size of a Cooper pair.'' Phase slips whose cores are
of that size are not seen in the standard GL theory (which treats the pairs as
pointlike) and thus represent an alternative channel of supercurrent decay.

A discrete model of superconductor has been used, 
for instance, in a study \cite{Matveev&al}
of the current-phase relation (CPR) of superconducting nanorings.
Various point of similarity
with the results of the GL theory have been noted.
The CPR, however, is primarily a low-current effect 
(unless, that is, one starts looking at higher bands in
the CPR band structure). In contrast, here
our focus is on the novel features that
a discrete model (similar but not identical to that in \cite{Matveev&al})
can bring in at currents close to the critical (depairing) current $I_c$.

At first glance, it seems easy to explain (\ref{3/2}) in a discrete model. 
For example, let the supercurrent through each link be
a sinusoidal function of
the phase difference
$\theta_{j+1} - \theta_j$ (where $j$ labels the nodes). The metastable 
ground state at a biasing current $I_b < I_c$ corresponds 
to all $\theta_{j+1} - \theta_j = \vgs$,
where $\vgs$ is the smaller root of
\be
\sin \vgs = \bmi_b
\label{sine_eq}
\ee
on the interval $[0,\pi]$. 
Consider now the configuration in which 
the phase difference on one of the links is instead the larger root of 
(\ref{sine_eq}), namely, $\pi - \vgs$. 
It seems reasonable to assume that this configuration is 
the ``critical droplet''---the saddle point 
whose energy determines the height of the activation 
barrier. Since, compared to the ground state, the phase difference changes on
only one link, that height is exactly the same as for a single junction and, in
particular, scales according to (\ref{3/2}).

A potential difficulty with this explanation becomes apparent if we observe that
the activation process envisioned above requires a change in the total phase
difference between the ends of the wire: if there is a total of $N$ links, this 
phase difference equals $N \vgs$ for the ground state and 
\[
\pi - \vgs + (N - 1) \vgs  > N \vgs 
\]
for the purported critical droplet. Consider a very long wire or a short wire connected to
bulk superconducting leads. In either case, one expects that the phases at the ends
will not be able to react
instantaneously to whatever changes occur in the middle; there will be a delay. 
In the limit of a large delay, the correct boundary conditions
are that the phases at the ends do not change at all during the
activation process and can only change during the subsequent real-time evolution.
(This is similar to how, in a first-order phase transition, 
a bubble of the new phase nucleates locally and then expands to fill the entire 
sample.) 

For a long wire, one may suspect that the change in the saddle-point energy brought
about by the boundary conditions is slight. To argue that more rigorously, one may allude
to the ``theorem of small increments'' \cite{LL:Stat}
(which relates the changes of different thermodynamic
potentials under small perturbations), as has been done by McCumber
\cite{McCumber} for the case of the LA saddle point in the GL theory.\footnote{Incidentally, 
for currents near the critical, this argument
(based on the premise that a localized nucleation event can be considered as a small 
perturbation)
appears more straightforward in the discrete model than in the original GL case. 
This is because in the former case the spatial size 
of the saddle point remains fixed in the limit $\bmi_b \to 1$, while in the latter 
it grows as $(1- \bmi_b)^{-1/4}$.} On the other hand, for a short wire (connected to 
bulk superconducting leads), there is no
reason to expect that the role of boundary conditions will be small. On general grounds,
one may expect that the activation barrier in this case will be higher than for a long 
wire, reflecting the tendency of superconductivity to be more robust in the presence of
the leads. The question, however, is whether this tendency will lead only to a larger value
of the constant in (\ref{3/2}), or it is capable of modifying the scaling exponent itself. 

In the present paper, we would like to answer this question. We would also like to 
understand how the transition from thermal activation to tunneling, as the main mechanism
of phase slips, occurs as the temperature is lowered. 
We describe the discrete model used in this paper in more detail in Sec.~\ref{sec:model}. 
We show (in Sec.~\ref{sec:static}) that, in this model,
the scaling law (\ref{3/2}) holds at least as long as the number $N$ of the links satisfies
$N > 4$,
despite the presence of boundary conditions
that suppress fluctuations at the ends. (Smaller values of $N$ constitute special cases, 
which we have not studied in detail.)

We then consider (in Secs.~\ref{sec:cross} and \ref{sec:comp}) the 
crossover to tunneling. In the semiclassical approximation, 
the rate of tunneling is determined
by classical solutions (instantons) that depend nontrivially on the Euclidean time
$\tau$ and are periodic in $\tau$ with period $\beta = \hbar / T$, where $T$ is the
temperature. We expect the semiclassical approximation to apply when 
the instanton action is 
large. Following \cite{periodic}, we consider solutions (periodic instantons) 
that have two turning points---states where all canonical momenta vanish 
simultaneously---one at $\tau = 0$ and the other at $\tau = \beta/2$. These states can
be interpreted as the initial and final states of tunneling. We search for periodic
instantons numerically and find that they 
exist at any temperature below a certain crossover temperature $T_q$. The latter 
scales near the critical current according to
\[
T_q = \mbox{const} \times (1 - \bmi_b)^{-1/4} \, .
\] 
At any $T < T_q$, the action of the periodic instanton is smaller than the activation
exponent $\Delta E / T$, which makes tunneling the main mechanism of phase slips at 
these temperatures.
At $T=0$, the instanton action in the discrete model scales at $\bmi_b \to 1$ as
\[
S_{inst}(T=0) = \mbox{const} \times (1 - \bmi_b)^{5/4} \, .
\]
Note that this scaling is different from the scaling law (\ref{3/2}) for the activation
barrier. The difference may be attributed to the ``critical slowing down'' of the 
Euclidean dynamics at currents near the critical.

\section{The discrete model} \label{sec:model}
We consider a finite chain of nodes, labeled by $j=0,\dots, N$, having
coordinates $x_j$ along a line in the physical space
and ordered according to $x_j < x_{j+1}$. 
On each node there is a phase variable, $\th_j$,
interpreted as the phase of the
superconducting order parameter at the corresponding point in the wire. We assume that
the nodes are equally spaced:
\[
x_{j+1} - x_j = \Delta x
\]
for all $j = 0, \dots, N-1$.

As we will see, solutions corresponding to phase slips in the present model involve
changes of $\th_j$ over a variety of spatial scales, including significant changes
on the scale $\Delta x$ (i.e., from one node to the next). Clearly, then,
in application to nanowires, results obtained with the help of this model
can be at best semiquantitative. Our main interest here is 
not so much in precise quantitative detail (although we will present estimates for
various quantities as we go) as in the scaling laws for rate
exponents at currents close to the critical. One may hope those to be
to some degree universal.

To estimate $\Delta x$ (and so also the number of nodes needed to describe
experimentally relevant wire lengths), we 
interpret the ratio $(\th_{j+1} - \th_j)/\Delta x$ as the gradient of the
phase of the order parameter, i.e. (up to a factor of $\hbar$), 
the momentum of a Cooper pair in the link
$(j,j+1)$. The value of this ratio corresponding to the maximum (critical)
current will be the critical momentum. To find it in the present model, we 
need an expression for the
supercurrent $I_{j,j+1}$ in the link as a function
of $\th_{j+1} - \th_j$. In the main text, we use the simplest $2\pi$-periodic
expression, the sinusoidal
\be
I_{j,j+1} = I_c \sin(\th_{j+1} - \th_j) \, ,
\label{cur}
\ee
but this choice is more or less arbitrary. As shown in the Appendix,
many results, including the scaling law (\ref{3/2}), generalize to 
a much broader class of $2\pi$-periodic functions. 

The choice (\ref{cur}), together with our later choice of the kinetic term 
for $\th_j$, 
makes our system equivalent to a particular model of a chain of Josephson 
junctions, the ``self-charging'' model of Ref.~\cite{Bradley&Doniach}, where
the dynamics of this model has been considered in the limit of large $N$ and zero
current. Note that, 
to describe a uniform wire, we have taken the critical current 
$I_c > 0$ to be the same for all the links. 

The current (\ref{cur}) reaches maximum when the phase difference equals $\pi / 2$,
so the critical momentum, as defined above, is
$p_c = \pi \hbar  / (2 \Delta x)$. If the wire were in the clean limit, we could 
estimate $\Delta x$ by comparing this to the expression from the microscopic theory,
$p_c = \Delta_0 / v_F$,
where $\Delta_0$ is the gap at $T=0$, and $v_F$ is the Fermi velocity. This
microscopic formula relates $p_c$ to the distance that an electron in a clean sample
would travel 
during the time $\hbar / \Delta_0$ (up to a numerical factor of order one,
that distance coincides with Pippard's coherence length $\xi_0$).
As nanowires are more 
appropriately described by the dirty limit, we replace that distance with the one that
an electron, now in the presence of disorder, will diffuse over the same timescale:
\[
\xi_D = \left( \frac{\hbar D}{\Delta_0} \right)^{1/2} \, ,
\]
where $D$ is the diffusion coefficient. We then obtain
\[
\Delta x = \frac{\pi \xi_D}{2} \, .
\]
Taking for estimates $D = 1.2 \times 10^{-4}$ m$^2$/s, as appropriate for amorphous
MoGe \cite{Bezryadin:book}, and $\Delta_0 = 0.76$ meV (corresponding to $T_c = 5$ K), 
we find $\xi_D = 10$ nm and $\Delta x = 16$ nm. Thus,
a 150 nm long wire corresponds to $N \sim 10$.

Note that the physical meaning of the distance $\xi_D$ is that of the ``size of a Cooper
pair.'' As such, $\xi_D$ is physically distinct from the GL coherence length $\xi_{GL}$
and, indeed, will not be even seen in the standard GL theory (which treats the pairs as
point-like). In this respect, a phase slip here, which takes place on the scale $\Delta x$,
and the one mediated by the LA saddle point \cite{LA} of the GL theory 
represent two different channels of
supercurrent decay. In this paper, we consider only temperatures significantly below
the critical and will not ask how the effective GL description at $T$ close to
$T_c$ arises in the discrete model.

Next, we formulate the equations of motion for the phases $\th_j$. We take the dynamics to
be entirely Lagrangian at the interior nodes 
but to have a dissipative component at the ends. The Lagrangian is
\be
L =  \half \sum_{j = 0}^N C_j (\partial_t \th_j)^2  - U \, ,
\label{L}
\ee
where the potential energy is
\be
U = - I_c \sum_{j=0}^{N-1} \cos (\th_{j+1} - \th_j) -  I_b (\th_N - \th_0) \, .
\label{pot_ene}
\ee
Here and in what follows, we use the system of units in which $\hbar$
and the charge of a pair are
set equal to 1: $\hbar = 1$, $2 e = 1$. Powers of $2e$ can be restored in the final
answers via the replacements 
\be
I_c \to I_c / 2e \, , \hspace{3em} C_j \to C_j / (2e)^2 \, .
\label{phys_units}
\ee
The kinetic term in (\ref{L}) corresponds to all nodes having finite capacitances, $C_j$,
to nearby conductors; these capacitances are assumed to the much larger than those
between the nodes. In this respect, the present model is different from that used in
\cite{Matveev&al}. 

In numerical computations, we will, for definiteness, assume
that all the $C_j$ are equal, except possibly for the two nodes at the ends.  
The static solutions, described in the next section, 
are insensitive to this assumption. The numerical method used to obtain 
the non-static solutions could be used also if the capacitances were distinct.

In (\ref{pot_ene}), the cosine term corresponds to our choice of the sinusoidal
expression (\ref{cur}) for
the current. The second, non-periodic term reflects the fact that we are considering the 
wire at a fixed biasing current, $I_b$. 
Formally, this term can be seen as a result of the Legendre 
transform with respect to $\phi = \th_N - \th_0$. Physically, it represents the work done by an external
battery to replenish the current back to the bias value. Note that the value of this term changes 
(i.e., the work done is nonzero) only when there is a change in $\phi$, the total phase accumulation
along the wire.

The equation of motion at the interior points is derived from the Lagrangian and reads
\be
C_j \ddot{\th}_j = I_c \sin (\th_{j+1} - \th_j) - I_c \sin (\th_j - \th_{j-1}) \, ,
\label{eqm}
\ee
$j=1,\dots,N-1$. At the endpoints, $j = 0$ and $N$, we include
dissipative dynamics intended primarily
to model suppression of quantum fluctuations by the bulk superconducting leads.
It also provides a mechanism for relaxation of the supercurrent back to the bias
value after thermal activation or tunneling.
We describe it by two small impedances, $R_0$ and $R_N$, 
shunting the ends of the wire to the ground. In classical theory, their effect
is represented by dissipative terms added to the equations of motion, as follows:
\ba
C_0 \ddot{\theta}_0 & = & I_c \sin(\theta_1 - \theta_0) - I_b - R_0^{-1} \dtheta_0 \, 
\label{eqm_0} \\
C_N \ddot{\theta}_N & = & - I_c \sin(\theta_N - \theta_{N-1}) + I_b - R_N^{-1} \dtheta_N \, .
\label{eqm_N}
\ea
The effect of the impedances {\em during} tunneling is represented by a 
Caldeira-Leggett term \cite{Caldeira&Leggett} in the Euclidean action. We do not
write this term explicitly here but merely assume that 
$R_0$ and $R_N$ are small enough for
it to suppress variations of the phase at the endpoints down to negligible
values.

\section{Static solutions} \label{sec:static}
For time-independent (static) solutions, the equations of motion become
\be
\sin (\th_{j+1} - \th_j) - \sin (\th_j - \th_{j-1}) = 0
\label{eqm_static}
\ee
at the interior points, and
\ba
\sin (\th_1 - \th_0) - \bmi_b & = & 0 \, , \label{sin_bc1} \\
- \sin (\th_N - \th_{N-1}) + \bmi_b & = & 0  \label{sin_bc2} 
\ea
at the ends. Here
\[
\bmi_b = I_b / I_c \, ,
\]
which without loss of generality can be assumed non-negative. Thus, it is 
in the range $0\leq \bmi_b \leq 1$.

In this section, we consider three types of solutions to these equations. The first type
is the ground states, one for each value of
\be
\vgs = \arcsin \bmi_b \, .
\label{vgs}
\ee
As we will see, these ground states are stable against small fluctuations
for all $0\leq \vgs < \pi/2$.
On the other hand, for any $\bmi_b > 0$ the potential (\ref{pot_ene}) is unbounded from
below, so these states are not absolutely stable but only metastable, i.e., 
subject to decay via large fluctuations, either
thermal or quantum. 

The other two types of solutions considered in this section correspond to certain
intermediate states in the decay of the (metastable) ground states. We view such a 
decay, whether it is classical or quantum, as a two-stage process, where at the first stage
the system reaches an intermediate state, which sits either at 
the top of the potential barrier (in the classical case) or on the other side of it (in
the quantum case). Crucially, we assume that the boundary values of $\th_j$ in this intermediate 
state are the same as they were in the ground state. 
This is consistent with our earlier discussion of how
the correct boundary conditions during the activation process must reflect 
the role of the superconducting leads. 
The current in such an intermediate state is different from the biasing current. At the second 
stage, the system slowly 
adjusts the boundary values of the phase to return the current back to the bias value.
Note that, since any resistive effect of a phase slip requires a change
in $\phi = \th_N - \th_0$, all such effects are relegated to the second stage.

In the two-stage picture, the configurations corresponding to the intermediate states need
to solve only the interior equations (\ref{eqm_static}) and not the boundary equations
(\ref{sin_bc1})--(\ref{sin_bc2}). Instead, the boundary values of the phase in these states
are determined
by the boundary conditions
\ba
\th_0 & = & 0 \, , \label{theta_bc1} \\
\th_N & = & \vgs N  \label{theta_bc2}
\ea
(the phase $\th_0$ can be arbitrarily set to zero, because the static equations involve
only the phase differences).

\begin{figure}
\begin{center}
\includegraphics[width=3.25in]{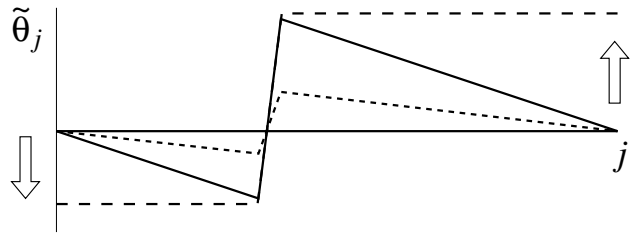}
\end{center}                                              
\caption{{\small Different configurations of the reduced phase (\ref{tth}) 
that solve the static equations of motion (\ref{eqm_static}) at the interior points. 
The ground state is $\tth_j \equiv 0$; it solves also the static equations
(\ref{sin_bc1})--(\ref{sin_bc2}) at the ends. Short dashes represent the saddle point
(critical droplet) that sits at the top of the potential barrier responsible for metastability of 
the ground state. The solid polyline is a state on the other side of the barrier; it is
referred to as a state with a $2\pi$ jump. Thermal decay corresponds to a transition
through a vicinity of the critical droplet, and tunneling {\em at low currents}
to transition from the ground state to the state with a $2\pi$ jump. 
(For tunneling at a high current, the final state needs to be found by the method described
in Sec.~\ref{sec:comp}.)
The state with a $2\pi$ jump does not solve the static equations at the ends, and so
will be subject to forces there. Those will induce its evolution, over a longer timescale,
towards the state shown by the long dashes (the initial direction of the evolution is
shown by the arrows).}
}                                              
\label{fig:gs}                                                                       
\end{figure}

Instead of the phase variable $\th_j$, we will often use the ``reduced'' variable $\tth_j$, 
with a linear growth subtracted away as follows:
\be
\tth_j = \th_j - \vgs j \, .
\label{tth}
\ee
For this, the boundary conditions (\ref{theta_bc1})--(\ref{theta_bc2}) become simply 
\be
\tth_0 = \tth_N = 0 \, .
\label{bc_tth}
\ee
An example of all three types of solutions (a ground state and the two
states mediating its decay) is shown in Fig.~\ref{fig:gs}.

For the analysis of linear stability of these various solutions, we will need 
the Hessian---the matrix of second derivatives of the potential energy (\ref{pot_ene}).
It is a symmetric tridiagonal matrix, which we prefer to write in units of $I_c$:
\be
M_{jj'} =  I_c^{-1} \frac{\partial^2 U}{\partial \th_j \partial \th_{j'}} \, .
\label{sec_var}
\ee
Consistent with the boundary conditions (\ref{theta_bc1})--(\ref{theta_bc2}), we restrict
variations of $\th_j$ to be supported at the interior points only; thus, in (\ref{sec_var}), 
$j,j' = 1, \dots, N-1$. Then, the diagonal elements of the Hessian are
\be
M_{jj} = \cos (\th_{j+1} - \th_j) + \cos (\th_j - \th_{j-1}) \, ,
\label{diag}
\ee
and the off-diagonal ones are
\be
M_{j,j+1} = M_{j+1,j} = - \cos(\th_{j+1} - \th_j) \, .
\label{off-diag}
\ee
We now consider the different types of solutions in turn.

\subsection{Ground states}
\label{subsec:ground}

The simplest solution to (\ref{eqm_static}) is the one where all the phase differences are
equal, i.e., $\th_j$ grows linearly with $j$ with some slope $\vgs$:
\be
\th_j = \vgs j \, .
\label{gs}
\ee
This solves also the boundary equations (\ref{sin_bc1})--(\ref{sin_bc2}) provided the slope
satisfies
\be
\sin \vgs = \bmi_b \, .
\label{sin_vgs}
\ee
As already done previously, for $\bmi_b < 1$,
we take $\vgs$ to be the smaller root of (\ref{sin_vgs}) 
on $[0,\pi]$. Eq.~(\ref{gs}) would remain a solution if we were to replace $\vgs$ with
the larger root, $\pi - \vgs$. As we will see, however, that latter solution is linearly
unstable. 

We refer to (\ref{gs}) as the ground state (corresponding to a given $\bmi_b$).
As noted earlier, the ground state is at best metastable (i.e., cannot
be absolutely stable) for any $\vgs > 0$.
In terms of the reduced phase (\ref{tth}), the ground state becomes simply 
\[
\tth_j \equiv 0 \, .
\]
Thus, in general, 
we can think of the reduced phase as measuring fluctuations relative to the ground
state.

Turning to analysis of linear stability, we note that all the cosines 
in the expressions (\ref{diag})--(\ref{off-diag})
are now equal to $\cos\vgs$. The eigenvectors of the Hessian are then readily found: they
are
\[
\alpha_j^{(p)} = \sin (p j) \, ,
\]
where $p$ is a positive integer multiple of $\pi / N$.
The corresponding eigenvalues are
\be
\lambda_p = 2 (1 - \cos p) \cos\vgs \, .
\label{lambda_gs}
\ee 
We see that the ground-state is linearly stable 
for $0 \leq  \vgs < \pi / 2$. The solution, obtained by replacing $\vgs$ with the
larger root of (\ref{sin_vgs}), i.e., $\pi -\vgs$, is linearly unstable. 
For the critical $\vgs = \pi/2$, the Hessian vanishes, and stability 
is determined by the leading non-linear term. That is cubic, and so the critical
solution is unstable.

\subsection{$2\pi$ jumps}
\label{subsec:jumps}
The equality of the sines in (\ref{eqm_static}) does not require equality of the arguments, 
and indeed (\ref{eqm_static}) has
solutions for which the phase differences across the links 
are not all equal. The simplest case is when
they are all equal except for one, which differs from the rest by $2\pi$. For the phase
itself, we then have
two segments of linear growth joined together by a jump of approximately $2\pi$
(cf. Fig.~\ref{fig:gs}).
In terms of the reduced phase (\ref{tth}), the solution satisfying the boundary 
conditions (\ref{bc_tth}) is 
\ba
\tth_j & = & -\frac{2\pi j}{N} \, , \hspace{5em} j = 0,\dots,k \, , \nonumber \\
\tth_j & = & -\frac{2\pi j}{N} + 2\pi \, , \hspace{3em} j = k + 1,\dots, N \, . \label{jump}
\ea
The jump occurs on the link from $j = k$ to $j = k + 1$, where $k$ can be any integer from
0 to $N-1$. 
We refer to it as a ``$2\pi$ jump'' even though the actual difference
\[
\tth_{k+1} - \tth_k = 2 \pi (1 - 1/N)
\]
is somewhat smaller than $2\pi$.

Analysis of linear stability is completely parallel to that for the ground state, with
stability now determined by the sign of 
$\cos(\vgs - 2\pi / N)$. For 
\be
N  >  4 \, , 
\label{N>4}
\ee
the solution is linearly stable for all $0\leq \vgs < \pi/2$.
Smaller values of $N$ constitute special cases. For $N =2$, 
the solution is unstable for any of the above $\vgs$. For $N = 3$, it is
unstable for $\vgs < \pi / 6$, while for $N = 4$ it is stable as long as $\vgs \neq 0$ 
with the
Hessian vanishing at $\vgs = 0$. In what follows, we do not pursue detailed analysis 
of these special cases but simply assume that the condition (\ref{N>4}) is satisfied.

What is the relevance of these solutions to decay of supercurrent? Our numerical 
results (a sample of which will be presented later) indicate that at temperatures
$T\to 0$ and {\em low currents}, 
the state with a $2\pi$ jump is the final state of the tunneling process, by which
the system, originally in a vicinity of the 
ground state, escapes through the potential barrier. The evolution after tunneling
(the second stage in our two-stage description) 
proceeds in real time, according to the equations of motion
(\ref{eqm}) and (\ref{eqm_0})--(\ref{eqm_N}) with $\tth_j$ = (\ref{jump}) and
$\partial_t \tth_j \equiv 0$ as the initial conditions. 
Note that (\ref{jump}) solves
the static equations at the interior points but not at the ends. Thus, the initial
direction of the evolution is determined by the net forces in the
boundary equations
(\ref{eqm_0})--(\ref{eqm_N}). The force at $j=0$ is equal to $I- I_b$, where
\[
I = I_c \sin (\vgs - 2\pi / N )
\]
is the current on the solution, and $I_b = I_c \sin \vgs$ is the biasing current;
the force at $j=N$ is $I_b - I$.
For $N > 4$ and 
$0\leq \vgs < \pi /2$, the difference $I - I_b$ is negative. This means 
that the initial direction of the real-time 
evolution is as indicated by arrows in Fig.~\ref{fig:gs},
that is the system evolves
towards the state shown in Fig.~\ref{fig:gs} by the long dashes. 

The state shown by the long dashes has $\tth_j = c$ (a constant) for $j \leq k$ and 
$\tth_j = c + 2\pi$ for $j \geq k +1$. 
Because each $\tth_j$ is defined modulo $2\pi$, this state differs from the ground
state $\tth_j \equiv 0$ essentially by a constant shift of the phase. 
The transition to it from
$\tth_j \equiv 0$, however, is observable as
it generates a voltage pulse in the external circuit and
involves work done by an external battery. This transition constitutes a phase slip.
The work done by the battery is given by the negative of the second, non-periodic
term in the potential energy (\ref{pot_ene}). For instance, for the case
shown in Fig.~\ref{fig:gs}, it equals $2\pi I_b$.

We wish to reiterate that this special role of (\ref{jump}) as the final state of
tunneling is characteristic of the low-current regime. At large currents, the final
states need to be found by the method described in Sec.~\ref{sec:comp}.

\subsection{Saddle points (critical droplets)}
One more way to solve the sine equation (\ref{eqm_static}) 
is to let all $\th_{j+1} - \th_j$ 
except one be equal, as in the case of a $2\pi$ jump,
but let that one be $\pi$ minus any other.
That is, if 
\be
\th_{j+1} - \th_j = \vgs + \gamma
\label{lin}
\ee
for $j \neq k$ (where $\gamma$ is an as yet undetermined constant), then
\be
\th_{k+1} - \th_k  = \pi - (\vgs + \gamma) \, .
\label{pi_minus}
\ee
In terms of the reduced phase (\ref{tth}), the solution corresponding to (\ref{lin}) and
satisfying the boundary conditions (\ref{theta_bc1})--(\ref{theta_bc2}) is
\ba
\tth_j & = & \gamma j \, , \hspace{5em} j = 0,\dots,k \, , \nonumber \\
\tth_j & = & \gamma (j - N) \, , \hspace{2em} j = k + 1,\dots, N \, . \label{drop}
\ea
Eq.~(\ref{pi_minus}) then becomes an equation for the slope $\gamma$.
It has a solution,
\be
\gamma = - \frac{\pi - 2 \vgs}{N - 2} 
\label{gamma}
\ee
for any $N > 2$. Note that $\gamma < 0$ for any $0\leq \vgs < \pi/2$ and is zero for the 
critical $\vgs = \pi/2$. In the latter case, the solution coincides
with the ground state.

Even though the solution exists for any $N > 2$, we restrict our attention to cases
when 
\be
\cos(\gamma + \vgs) = \cos \frac{N \vgs - \pi}{N - 2}  > 0 \, ,
\label{cos}
\ee
because then the solution has a special interpretation, discussed below. 
This inequality is always satisfied
for $\vgs < \pi/2$ and $N > 4$, a condition we have already imposed.

We will refer to the link form $j=k$ to $k+1$ as the ``core'' of the solution, even 
though, as we will see shortly, the energy is by no means concentrated at the core:
the difference in the slope from the ground state is essential and leads to 
a contribution
distributed over the entire length of the wire.

The solution is shown in Fig.~\ref{fig:gs} by the short dashes. The way it appears
there implies
that the slope $|\gamma|$ is smaller than $2\pi / N$, the slope of the solution with
a $2\pi$ jump. One can verify that the condition for that is precisely the same as
the condition of linear stability of the latter solution, derived previously. 

Given that, as
far as the slopes go, the present solution lies between the ground state and the state
with a $2\pi$ jump, one may expect that it is a saddle point that sits at the top of
the potential barrier separating the two states. We will now see that it is in fact
a {\em critical droplet}---a saddle point that has exactly one negative mode and 
mediates thermally activated decay of the ground state. 

A negative mode is an eigenvector of the Hessian corresponding to a negative eigenvalue.
The cosines in eqs.~(\ref{diag})--(\ref{off-diag}) now are
\[
\cos(\th_{j+1} - \th_j) = \left\{ \begin{array}{cc} 
\cos(\vgs + \gamma) \, , & j\neq k \, , \\
-\cos(\vgs + \gamma) \, , & j = k \, .
\end{array} \right.
\]
Thus, the eigenvalue problem for the Hessian is a discrete version of one-dimensional 
quantum mechanics with a localized potential.
A negative mode corresponds to a bound state. 
The corresponding eigenvalue is of the form
\be
\lambda_- = - \Lambda \cos(\vgs + \gamma) \, ,
\label{lam_neg}
\ee
where $\Lambda > 0$ depends only on $N$ and $k$. 

For given $N$ and $k$, it is straightforward to find eigenvalues of the Hessian by 
numerical diagonalization. We have done that for a few small values of $N$, to convince
ourselves that there is a unique negative mode in those cases. For $N \gg 1$,
it is possible to prove existence and uniqueness of a negative mode
without resorting to numerics.

As a sample of these
results, consider the case when $N$ is odd and $k = \half (N - 1)$, meaning that the
core of the solution is directly in the middle of the wire. 
We find, for instance, 
$\Lambda = 1.303$ for $N = 5$ and $\Lambda = 1.330$ for $N = 7$. The latter value is
already close to the asymptotic
\be
\Lambda = \frac{4}{3} 
\label{Lam}
\ee
which corresponds to $N\to \infty$ and can be found analytically. 
The eigenvector corresponding to the negative mode is
\be
\aneg_j = \left\{ \begin{array}{cc} 
A \sinh (p j) \, , & j\leq \half (N -1) \, , \\
- A \sinh [p (N - j)] \, , & j\geq  \half (N + 1) \, . \end{array} \right.
\label{neg_mode}
\ee
where $A$ is a normalization coefficient, and $p$ is related to the eigenvalue by
\[
\Lambda = 2 \cosh p - 2 \, .
\]
For $N \to \infty$, $p = \ln 3$.
Moving the core away from the middle causes $\Lambda$ to decrease, but the negative 
mode persists for all $k$, even $k = 0$. In the latter case, the asymptotic value 
of at $N \to \infty$ is $p = \ln 2$, that is $\Lambda = \half$. 

Note that, unlike the droplet itself,
which has linear ``tails'' extending to the ends
of the wire (cf. Fig.~\ref{fig:gs}), the negative mode is tightly localized: 
for instance, $p = \ln 3$ (in the $N\to\infty$ case) means that the 
magnitude of (\ref{neg_mode}) decreases by a factor of 3 per link
as one moves away from the core. As a consequence, the asymptotic result 
(\ref{Lam}), obtained for a droplet in the middle of a long
wire, has exponential accuracy in $N$ and will hold well even for droplets
away from the middle, except for very short wires or droplets very near the ends.

When the droplet is strictly in the middle, 
the negative mode is antisymmetric about
the core. This will hold approximately also away from the middle, except when
the droplet is close to one of the ends.
This antisymmetry has a simple interpretation: as clear from Fig.~\ref{fig:gs},
adding an antisymmetric 
$\aneg_j$ with a small coefficient (i.e., moving along the negative mode)
will deform the droplet either towards the
ground state or towards the state with a $2\pi$ jump,
precisely as expected 
of motion across the top of the potential barrier separating the two states.

\subsection{Activation barrier}
\label{subsec:act}

The energy of the critical droplet is obtained by substituting (\ref{lin}) and
(\ref{pi_minus}) into (\ref{pot_ene}). 
In units of $I_c$, it equals
\be
\edrop \equiv E_{drop} / I_c =
- (N - 2) \cos(\gamma + \vgs) -  \bmi_b (\th_N - \th_0) \, .
\label{Edrop}
\ee
This should be compared to the energy of the ground state,
\be
\egs \equiv  E_{gs} / I_c = - N \cos \vgs -  \bmi_b (\th_N - \th_0) \, .
\label{Egs}
\ee
The difference between the two is the activation barrier for thermally activated phase 
slips (TAPS). Using (\ref{cos}), we obtain
\be
\edrop - \egs = N \cos \vgs - (N - 2) \cos \frac{N \vgs - \pi}{N - 2} \, .
\label{Eact}
\ee
A curious property of (\ref{Eact}) is that it is independent of $k$, the
location of the droplet core. In a continuum model, that would imply existence of
a zero mode---a zero eigenvalue of the Hessian---associated with the translational 
symmetry. In the
present case, there is no such mode, as the symmetry with respect to infinitesimal
translations is broken by the lattice.

Another convenient expression for the activation energy is obtained by using,
instead of $\vgs$,  the parameter $\eps$ defined by
\be
\vgs = \frac{\pi}{2} - \eps \, .
\label{eps}
\ee
Then,
\be
\edrop - \egs = N \sin\eps - (N - 2) \sin \frac{N \eps}{N-2} \, .
\label{Eact_eps}
\ee

Eq.~(\ref{Eact}) has two interesting limits. The first is $N \gg 1$ with $\vgs$ fixed.
In this case, 
\be
\edrop - \egs = 2 \cos \vgs - (\pi - 2 \vgs) \sin \vgs + O(1/N) \, .
\label{E_lim1}
\ee
We see that the activation barrier remains finite in the limit of large length.
That does not
mean, however, that the energy difference (\ref{E_lim1}) is accumulated locally, around 
the core of the droplet: the reduction (by the amount $|\gamma|$) of the slope of 
the droplet's ``tails'' relative to the ground state, cf. (\ref{lin}), is essential
and leads to a contribution distributed over the entire length of the wire.

Expressing $\vgs$ through the biasing current via (\ref{sin_vgs}), we find that,
in the limit $N\to \infty$, the activation barrier (\ref{E_lim1}) coincides exactly
with that of a Josephson junction whose potential energy, as a function of the
phase difference, is
\be
U(\phi) = - I_c \cos\phi - I_b \phi \, .
\label{UJJ}
\ee
As we show in the Appendix, this agreement is not limited to sinusoidal currents
but extends to a much broader class of current-phase relations.
It may not be obvious a priori, given especially that, in the wire, the activation process
does not involve any changes of the phase difference between the ends, so in the computation
of the activation barrier
the last terms in (\ref{Edrop})
and (\ref{Egs}) simply cancel each other. In contrast, in the junction (where the ground state
is $\phi_{gs}= \arcsin \bmi_b$, and the critical
droplet is $\phi_{drop} = \pi - \phi_{gs}$),
the contribution of the last term in (\ref{UJJ}) is
essential. Nevertheless, in the $N\to \infty$ limit, the agreement between the two cases
is not entirely
unexpected: it can be seen as an instance of the ``theorem of small
increments'' \cite{LL:Stat}, as already noted by McCumber \cite{McCumber} in the context of the
continuous GL theory. 

The second interesting limit of (\ref{Eact}) is $\vgs \to \pi / 2$ with $N$ fixed,
which corresponds to currents close to the critical. In this case, it is convenient to
use the form (\ref{Eact_eps}) of the activation energy: 
the parameter $\eps$ is now small, and we can expand (\ref{Eact_eps}) in it. We obtain
\be
\edrop - \egs = \frac{2 N (N-1)\eps^3}{3(N-2)^2} + O(\eps^5) \, .
\ee
In terms of the biasing current,
\be
\eps = \sqrt{2} (1 - \bmi_b)^{1/2}  + O[(1 - \bmi_b)^{3/2}] \, .
\label{eps_scaling}
\ee
We see that the activation barrier is higher for smaller $N$, as might be expected, but
the scaling at $\bmi_b \to 1$ is always the same 
$(1 -\bmi_b)^{3/2}$. This is the first of the results we have highlighted in the Introduction.

Finally, we note that, for $N > 4$, 
there is also a double droplet, a state where eq.~(\ref{pi_minus}) 
(with a different value of $\gamma$) is used on two links. Its energy,
in the same notation as in (\ref{Eact_eps}), is
\be
\edouble - \egs = N \sin\eps - (N - 4) \sin \frac{N \eps}{N-4} 
\ee
and is always larger than the energy of the single droplet.

\section{Euclidean solutions and the crossover temperature}
\label{sec:cross}
There can be no thermal activation at $T=0$, only quantum tunneling, so as the temperature
is lowered past some point one mechanism of phase slips must give precedence to the other.
In the semiclassical approximation,
tunneling is described by classical solutions (instantons)
that depend nontrivially on the Euclidean 
time $\tau = it$ and are periodic in $\tau$  with period $\beta = 1/T$.
Following \cite{periodic}, we consider ``periodic instantons''---solutions with two turning 
points (those are configurations where all the canonical momenta vanish simultaneously),
one at $\tau = 0$, and the other at half the period, $\tau = \half \beta$.
In the present
case, the turning-point conditions are
\be
\partial_\tau \tth_j(0) = \partial_\tau \tth_j(\beta /2) = 0 \, .
\label{bc_tau}
\ee
As discussed in \cite{periodic}, a periodic instanton 
is expected to saturate the microcanonical tunneling rate $\Gamma_{micro}$,
i.e., to give the most
probable tunneling path at a fixed energy $E$; the period of the instanton in that case
is determined by the energy. The same instanton will then saturate also the canonical
(fixed temperature) 
tunneling rate 
\be
\Gamma_{can}(\beta) \sim \int dE e^{-\beta E} \Gamma_{micro}(E) \, ,
\label{can}
\ee
provided the integrand here is peaked about the corresponding energy.
As we discuss in more detail below, such is indeed the case in our present
model.

The turning-point conditions (\ref{bc_tau}),
together with the spatial boundary conditions (\ref{bc_tth}), define a boundary problem
for the rectangle $0 < j < N$, 
$0\leq \tau \leq \half \beta$.
The solution for $\half \beta < \tau \leq \beta$ is obtained via
\[
\tth_j(\tau) = \tth_j(\beta - \tau) \, .
\]
The turning points $\tth_j(0)$ and $\tth_j(\beta/2)$ correspond, respectively, to the
initial and final states of tunneling, that is, the states in which the 
system enters and leaves the classically forbidden region of the configuration space.
The subsequent real-time evolution (the second stage in our two-stage description) 
proceeds with $\tth_j(\beta/2)$ as the initial state.
Thus, $\tth_j(\beta/2)$ must be real. Since, for $0 < j < N$, 
the Euclidean equations of motion contain no complex coefficients, the entire $\tth_j(\tau)$
must then be real, so we concentrate on real-valued solutions in what follows.

As shown in \cite{periodic}, the main, exponential factor in the microcanonical 
tunneling rate is
determined by the instanton's ``abbreviated'' (Maupertuis) action per period, $\tS(E)$,
as follows:
\be
\Gamma_{micro}(E) \sim e^{-\tS(E)} \, .
\label{micro}
\ee
So, the conditions for the integrand in (\ref{can}) to have a maximum at $E$ are
\ba
-d\tS / dE & = & \beta \, , \label{dtS} \\
d^2 \tS / dE^2 & > & 0 \, .  \label{d2tS}
\ea
If these are satisfied, 
\be
\Gamma_{can}(\beta) \sim e^{-S_{inst}(\beta)} \, ,
\label{can_exp}
\ee
where
\be
S_{inst}(\beta) = \beta E + \tS(E) \, .
\label{Sinst}
\ee
Note that the normalization of (\ref{can}) assumes that the ground state energy is zero; 
thus, $E$ in (\ref{Sinst}) is the energy of either turning point (by the
Euclidean energy conservation, their energies are equal) relative to the ground
state. Also, it is understood that $E$ is expressed through $\beta$ by means of (\ref{dtS}).
One result of that is the relation \cite{periodic}
\be
dS_{inst}/d\beta = E \, .
\label{dSdb}
\ee

By a standard theorem of mechanics \cite{LL:Mech}
(translated to the Euclidean time), the right-hand side of (\ref{dtS}) is the period of the 
solution, so (\ref{dtS}) tells us that the period is the same as $\beta$. This is not
unexpected but does go to show formally 
that (\ref{Sinst}) is the full (non-``abbreviated'') action 
of the instanton, relative to that of the ground state.
Note that the exponential factor (\ref{can_exp}) accounts both for the suppression of 
the rate due to the tunneling per se (as represented by $\tS$) and for that caused by 
the need to populate an initial state of energy $E$.

The inequality (\ref{d2tS}) is less trivial, in the sense that it may or may not be
satisfied in a given system for a given range of energies.
At small $E$, if the zero-energy instanton is known, one can construct an approximate 
periodic instanton
by alternating instantons and anti-instantons in the $\tau$ direction 
\cite{periodic} and check the condition (\ref{d2tS}) for it. 
For larger $E$, however, one typically has to resort
to numerical studies. For the latter, the condition (\ref{d2tS}) 
can be somewhat more conveniently rewritten as
\be
\frac{d^2 S_{inst}}{d\beta^2}  < 0 \, .
\label{d2S}
\ee
(By virtue of the relation between $E$ and $\beta$, the left-hand side here is 
the negative of 
$(d^2 \tS / d E^2)^{-1}$.)
Cases of varying complexity in the behavior of $\Sinst(\beta)$ have 
been described in the literature (as briefly surveyed below), 
and we now discuss the specifics of the present case.

In general,
the critical droplet undergoes a bifurcation at some $\beta = \bbif$, where it develops, 
in addition to its single $\tau$-independent negative mode, another, $\tau$-dependent one.
The key distinction between different cases is in where that new negative mode 
leads at $\beta$ just above $\bbif$: it can lead
to a limit cycle (a periodic instanton) in a vicinity of the critical droplet or, alternatively,
to an entirely different region of the configuration space. This distinction is analogous
to the one between the supercritical and subcritical forms of the usual Hopf bifurcation.
For the case at hand, we find
(numerically) that the former case (a nearby limit cycle)
is realized. In addition, we find that a real-valued  periodic instanton exists only
for $\beta > \bbif$, and for all these $\beta$ 
the condition (\ref{d2S}) is satisfied. 
In these respects, the present system is similar to the 
2-dimensional Abelian Higgs model; the periodic instanton for that case was found 
numerically in 
\cite{Matveev}. The behavior is different from that in the various versions of the 
2-dimensional $O(3)$ sigma model or in the 4-dimensional SU(2) Yang-Mills-Higgs theory; 
numerical solutions for those cases were found, respectively, 
in \cite{Habib&al,Kuznetsov&Tinyakov} and 
\cite{Frost&Yaffe:1998,Frost&Yaffe:1999,Bonini&al}.

At the bifurcation point, the condition (\ref{dSdb}) (recall that $E$ in it is measured from 
the ground state) implies 
\be
 \frac{dS_{inst}}{d\beta} (\bbif) = E_{drop} - E_{gs} \, .
\label{slope}
\ee
The right-hand side here is the activation barrier of the preceding section.
In other words, at this point, the $\Sinst(\beta)$ curve touches the straight line
\be
S_{drop}(\beta) = (E_{drop} - E_{gs}) \beta \, ,
\label{Sdrop}
\ee
corresponding to the thermal activation exponent. 
The condition (\ref{d2S}) being satisfied for
all $\beta > \bbif$ then implies that $\Sinst$ deviates down
from (\ref{Sdrop}), so that, at least in the leading semiclassical approximation 
(where only the exponential factors count), tunneling has a larger rate than thermal 
activation. Thus, in the present case, $\bbif$ is the same as $\beta_q$, the point
where one mechanism of phase slips overtakes the other.
Note that the crossover temperature $T_q = 1/ \beta_q$
can be measured experimentally 
\cite{Li&al,Aref&al,Belkin&al}.

Computation of $\bbif$ and so, in the present case, also of $T_q$ is standard. For 
the Lagrangian (\ref{L}), the $\tau$-dependent normal modes are
eigenfunctions of the operator
\be
\hat{N}_{ij} = - C_i \delta_{ij} \partial_\tau^2 + I_c M_{ij} \, ,
\label{Nij}
\ee
where $i,j = 1,\dots,N-1$, and $M_{ij}$ is the Hessian (\ref{sec_var}). As already mentioned,
in this paper, we set all the capacitances $C_j$ at the interior
points equal, $C_j = C$. Then, the eigenfunctions of (\ref{Nij}) are of the form
\[
\psi_j^{(n)}(\tau) = \alpha_j \cos(2\pi T n \tau) \, ,
\]
where $\alpha_j$ is an eigenvector of the Hessian, and $n \geq 0$ is an integer. The choice
of the cosine, rather than sine, here respects the boundary conditions (\ref{bc_tau}). 
In this way, for each eigenvalue $\lambda$ of the Hessian, the $\tau$-dependent problem
generates an infinite sequence of eigenvalues:
\be
\lambda \to \lambda + \left( \frac{2\pi T n}{\omega_c} \right)^2 \, ,
\label{time-dep}
\ee
where
\be
\omega_c = (I_c / C)^{1/2} \to (2e I_c / C)^{1/2} \, .
\label{omega_c}
\ee
In the last relation,
the arrow signifies transition to the physical units via (\ref{phys_units}).
For $\lambda > 0$, (\ref{time-dep}) can never produce a negative eigenvalue.
For $\lambda = \lambda_- < 0$, there is always the original one, corresponding to
$n = 0$. Another one appears at
\be
T < T_q = \frac{\omega_c |\lambda_-|^{1/2}}{2\pi} \, ,
\label{T_q}
\ee
which determines the crossover temperature. 

Recall that $\lambda_-$ depends not
only on $N$ and the biasing current $\bmi_b$, 
which are the parameters of the system, but also
on $k$, the location of the droplet along the wire. The meaning of $T_q$ is that of
the highest temperature for which tunneling is the main mechanism of phase slips, so
in (\ref{T_q}) we choose the largest $|\lambda_-|$ we can get at given $N$ and $\bmi_b$. 
As we have seen in the preceding section, this corresponds
to $k$ in the middle of the wire. With this choice, $\lambda_-$ is given by 
(\ref{lam_neg}) where $\Lambda$ now depends only on $N$.

Because $T_q$ can be measured 
experimentally, (\ref{T_q}) can be used to estimate
$\omega_c$. Let us use for this estimate
the asymptotic large-$N$ value $\Lambda = 4/3$,
which, as we have seen, becomes a good approximation already at modest $N$.
Setting $N= 11$ and the biasing current 
to 90\% of the critical current
results in $\omega_c = 7.5~T_q$. Since $I_c$ is also measurable,
we can use (\ref{omega_c}) to convert this estimate of $\omega_c$ into an estimate of $C$.
For $I_c = 10~\mu\mbox{A}$ and $T_q =0.9~\mbox{K}$ (corresponding to the values in the
experiment of \cite{Belkin&al}), we obtain
$C = 3.9 \times 10^{-14}~\mbox{F}$. Note that this is the capacitance of 
a single segment of length $\Delta x$, not of the entire wire. We attribute this
relatively large value of $C$ to the wire being in close proximity to large 
conductors (e.g., the center conductor strips \cite{Belkin&al}).

Next, consider the scaling of various quantities near the critical current.
The Hessian matrix of the critical droplet is proportional to $\cos(\gamma + \vgs)$.
In view of (\ref{cos}), this means that 
all its eigenvalues, including $\lambda_-$, scale at $\bmi_b \to 1$ as the first
power of $\eps= \pi/2-\vgs$, i.e., as $(1- \bmi_b)^{1/2}$. 
Then, according to (\ref{T_q}), $T_q$ scales
as $(1 -\bmi_b)^{1/4}$. 
This implies that the system stays classical longer (i.e., until a lower temperature)
as the current gets closer to the critical. The power-$1/4$ dependence, though, 
is quite weak,
and it is not clear if this effect can be observed experimentally.

The Hessian of the ground state is proportional to $\cos\vgs = \sin\eps$ and so also 
scales as $\eps$. Thus, at $T \ll T_q$ the characteristic frequencies of Euclidean motion 
near the ground state and near the top of the barrier are both of order 
$\omega_0 \sim \omega_c \sqrt{\eps}$.
This suggests that, at these $T$, we can estimate the instanton action $\Sinst$
by taking the product of the characteristic barrier height,
$I_c (E_{sph}- E_{gs})$, and the common timescale $2\pi / \omega_0$. The result is
\be
S_{inst} \sim 2\pi (I_c C)^{1/2} (1 - \bmi_b)^{5/4} \, .
\label{Sinst_est}
\ee
We will see that the power-$5/4$ scaling law for $\Sinst$ is well borne out numerically.
Note that, due to the ``critical slowing down'' of the Euclidean dynamics
at $\bmi_b \to 1$, this scaling law is different from the one for the barrier height
itself. This is the second of the results highlighted in the
introduction.

\section{Computation of the tunneling exponent}
\label{sec:comp}
We now turn to a systematic study of periodic instantons in our system.
The Euclidean equations of motion at the interior points, with all $C_j$ set
equal to $C$, read
\be
\partial_\tau^2 \tth_j + \omega_c^2 \left[ \sin(\tth_{j+1} - \tth_j + \vgs) 
- \sin(\tth_j - \tth_{j-1} + \vgs) \right] = 0 \, ,
\label{eqm_eucl}
\ee
where $\omega_c$ is the frequency (\ref{omega_c}). Recall that these
are to be solved on the rectangle $0 < j < N$, $0\leq \tau \leq \half \beta$
with the boundary conditions (\ref{bc_tth}) and (\ref{bc_tau}). Note that
this boundary problem depends only on the following dimensionless parameters:
$N$, $\vgs$, and $\omega_c \beta$.

The Euclidean action corresponding to this boundary problem is 
\be
S_E = \int_0^\beta \!\! d\tau  \left[ \frac{C}{2} 
\sum_{j=1}^{N-1}  (\partial_\tau \tth_j)^2 
- I_c \sum_{j=0}^{N-1} \cos(\tth_{j+1} - \tth_j + \vgs)  \right] . 
\label{SE}
\ee
The difference between (\ref{SE}) computed for the instanton and that for the
ground state $\tth_j \equiv 0$ gives the instanton action $\Sinst$ of the preceding
section:
\be
S_{inst} = S_E + \beta N I_c \cos\vgs \, .
\label{diff}
\ee
Under the conditions stated there, $\Sinst$ determines the exponential factor in the
tunneling rate at temperature $T = 1/\beta$, cf. eq.~(\ref{can_exp}).

In the preceding, we concentrated on the dependence of $\Sinst$ on the inverse 
temperature $\beta$. More generally, after extracting an overall factor, we
can write is as a function of the three dimensionless parameters
mentioned earlier and, in addition, of the instanton's spatial location; the latter
is labeled,
as in the case of the critical droplet,
by an integer $k$:
\be
S_{inst} = (I_c C)^{1/2} \sigma_{inst}(N, \vgs, \omega_c \beta; k) \, .
\label{resc_action}
\ee
The function $\sigma_{inst}$ will be referred to as the {\em reduced action}.
In what follows, we present results for it for the case when the 
instanton is in the middle of the wire; for odd
$N$, this corresponds to $k = \half (N - 1)$. We have also found solutions, albeit with
larger actions, with cores away from the middle.

Note that the activation exponent (\ref{Sdrop}) can be written, similarly to
(\ref{resc_action}), as
\[
S_{drop} = (I_c C)^{1/2} \sigma_{drop} (N, \vgs, \omega_c \beta) \, ,
\]
where
\be
\sigma_{drop} (N, \vgs, \omega_c \beta) = (\edrop - \egs) \omega_c \beta 
\label{resc_drop}
\ee
and,
as in Sec.~\ref{sec:static}, $\e$ denotes an energy measured in units of
$I_c$.
Thus, a comparison of $S_{inst}$ to $S_{drop}$ amounts to a comparison of
$\sigma_{inst}$ to $\sigma_{drop}$.

For numerical work, we discretize eq.~(\ref{eqm_eucl}) on a uniform grid in the
$\tau$ direction and solve the resulting difference equation by the Newton-Raphson 
method. This is the same general approach as used, for instance, 
in \cite{Frost&Yaffe:1999,Bonini&al}
to find periodic instantons in the Yang-Mills-Higgs theory.

\subsection{Tunneling at low currents}
\label{subsec:low}
We take up
the case of high biasing currents in the next subsection. Here, we briefly discuss the case
\be 
\vgs \leq \pi / N \, ,
\label{low_cur}
\ee
which we refer to as low currents. 

The condition (\ref{low_cur}) is a result of 
comparing the energy (\ref{Egs}) of the ground state to the energy of the state with
a $2\pi$ jump, eq.~(\ref{jump}). When (\ref{low_cur})
is satisfied,
\be
E_{jump} - E_{gs} = N [ \cos\vgs - \cos(\vgs - 2\pi / N) ] > 0 \, .
\label{up}
\ee
The linear stability analysis in subsec.~\ref{subsec:jumps} and the 
discussion there of the real-time evolution of the state with a $2\pi$ jump suggest
that, for $N > 4$,
this state is the lowest-energy state in which the system can emerge after tunneling
through the potential barrier. We have not rigorously proven
this assertion, but it is consistent with the numerical results, and in this
subsection we will
proceed on the premise that it is correct. Then, (\ref{up}) implies that, at low 
currents, a phase slip requires tunneling ``up'' in energy or, more precisely, that the
initial state of tunneling cannot be the ground state but must be thermally activated.
This means that (at low currents) the instanton action $\Sinst$ retains a nontrivial dependence
on $\beta$ for arbitrarily large $\beta$ (i.e., low temperatures), namely, that in this limit
\be
S_{inst}(\beta) \approx (E_{jump} - E_{gs}) \beta \, .
\label{Sjump}
\ee
The slope here, given
by (\ref{up}), is not as large as the full barrier height, eq.~(\ref{Eact}), because
the system does not need to activate to the top of the barrier but only to an initial
state with enough energy to tunnel to the state with a $2\pi$ jump. This activated
behavior can be traced back to the boundary conditions (\ref{bc_tth}) the phase
must satisfy during tunneling and so is ultimately
an effect of the bulk superconducting leads. As such,
it was identified in a different (continuum) model of the wire 
in \cite{Khlebnikov:CG}.

\begin{figure}
\begin{center}
\includegraphics[width=3.25in]{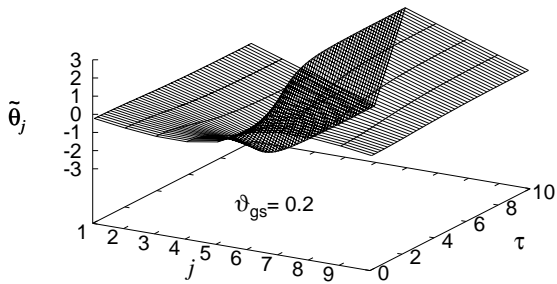}
\end{center}                                              
\caption{\small A low-current periodic instanton. The plot shows
the reduced phase $\tth_j(\tau)$, which is equal to
the difference between the original phase
variable $\th_j(\tau)$ and the ground state $\th_j = \vgs j$. 
The values of the  parameters are
$N = 11$, $\vgs = 0.2$, $\omega_c \beta = 20$, and $k = 5$. Recall that
$\sin\vgs$ is the biasing current in units of the critical. Only the
interior points $j = 1,\dots,10$ and half of the period $[0,\half \beta]$ in the $\tau$
direction are shown; $\tau$ is in units of $\omega_c^{-1}$. 
}                                              
\label{fig:small}                                                                       
\end{figure}

An example of numerically found low-current periodic instanton 
is shown in Fig.~\ref{fig:small}. The front of the plot ($\tau = 0$)
corresponds to the initial state
of tunneling, and the back ($\tau = \half \beta$) to the final state. The difference
between the initial state and the ground state $\tth_j \equiv 0$ represents the
thermal excitation required by the argument above. The final state coincides,
as far as we can tell,
with the state with a $2\pi$ jump.

\subsection{Tunneling at arbitrary currents}
\label{subsec:arb}

\begin{figure}
\begin{center}
\includegraphics[width=3.25in]{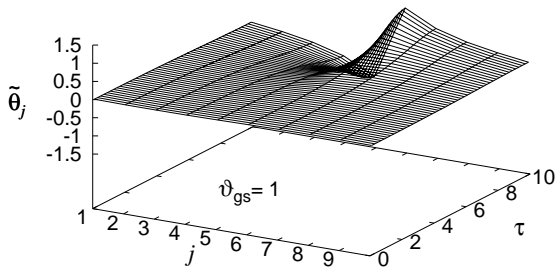}
\end{center}                                              
\caption{\small A high-current periodic instanton. The reduced phase $\tth_j(\tau)$
  for the instanton with the same
  $N$, $\omega_c \beta$, and $k$ as in Fig.~\ref{fig:small},
  but with a larger $\vgs = 1$.
}                                              
\label{fig:large}                                                                       
\end{figure}

A sample result for $\tth_j(\tau)$ for a comparatively large current, is shown
in Fig.~\ref{fig:large}. The key distinction between this instanton and the one for
small current, Fig.~\ref{fig:small}, is that now the system tunnels practically from
the ground state. That is so, even though the temperature for this plot
is only about a factor of 3 smaller
than the crossover $T_q$. In the final state, the phase still has
a characteristic jump at the tunneling location, but the magnitude of the jump 
is reduced compared to the low-current case.

\begin{figure}
\begin{center}
\includegraphics[width=3.25in]{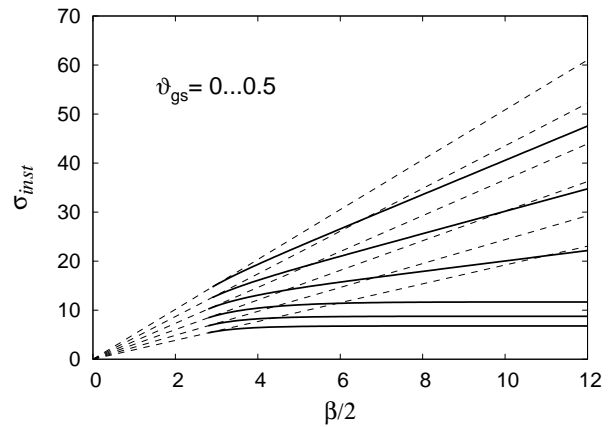}
\end{center}                                              
\caption{\small 
Solid lines: the reduced instanton actions $\sigma_{inst}$
for $N = 11$, $k =5$ (instanton in the middle of the wire), and several values of
$\vgs$. These are
plotted as functions of the half-period $\half \beta$, where $\beta$ is units of 
$\omega_c^{-1}$. The values of $\vgs$ increase in 
increments of 0.1 from top to bottom.
Dashed lines: the straight lines (\ref{resc_drop}) representing the over-barrier activation
exponents for the same $\vgs$.
The three smaller values
of $\vgs$ correspond to the low-current regime (cf. subsec.~\ref{subsec:low}), where
the instanton action displays the activated (linear) behavior (\ref{Sjump}) at large $\beta$
(though with a slope reduced compared to the full barrier height). 
}                                              
\label{fig:action}                                                                       
\end{figure}

In Fig.~\ref{fig:action}, we plot the reduced action $\sigma_{inst}$, 
as defined by (\ref{resc_action}), as a function of the half-period for several 
values of $\vgs$. These plots
illustrate the crossover from the
high-temperature (small $\beta$) 
regime, where the main mechanism of phase slips is thermal activation,
to the low-temperature (large $\beta$) regime, where the main mechanism is tunneling.
We see that, after the instanton first appears at $\beta= \beta_q$, its action 
for all $\beta > \beta_q$
lies below the  
straight line representing thermal activation---the
behavior announced in Sec.~\ref{sec:cross}. The crossover 
temperatures for all the curves shown are fairly close to one another 
and correspond to $\half \omega_c \beta_q \approx 3$. For lower values of the current,
the crossover leads to a transition from the high-temperature activated behavior to
one with a smaller slope, as anticipated in subsec.~\ref{subsec:low} 
(and found for a continuum model in \cite{Khlebnikov:CG}). 
For larger currents, the action quickly saturates 
at the zero-temperature value.

Next, consider dependence of the results on $N$, the length of the wire. 
The low-current condition (\ref{low_cur}) is obviously
sensitive to $N$. The results at higher currents, however, are not particularly so.
For instance, for $\vgs \geq 1$, doubling the length from $N=11$ to $N=21$ makes the
actions smaller by only a few percent.

Finally, let us discuss the case of currents close to the critical, $\vgs \to \pi/2$.
In this regime, we are mostly interested in the scaling law obeyed by the
asymptotic value of $\sigma_{inst}$ at $\beta \to \infty$,
as a function of $\eps = \pi /2 - \vgs$. We find that 
the power-$5/4$ scaling anticipated in (\ref{Sinst_est}) is well borne out. 
Numerically, for $N =11$,
\be
\sigma_{inst}(\beta\to \infty) = 5.5 ~\eps^{5/2} 
\label{sigma}
\ee
at $\eps \to 0$; for $N=21$, the coefficient changes from 5.5 to 5.2.
Substituting (\ref{eps_scaling}) for $\eps$, we find that (\ref{sigma}) corresponds to 
\be
S_{inst}(T\to 0) \approx 13 \left( \frac{\hbar I_c C}{8 e^3} \right)^{1/2} (1 - \bmi_b)^{5/4} \, ,
\label{T=0}
\ee
where we have also used (\ref{phys_units}) to convert $(I_c C)^{1/2}$ to the physical units.
For $I_c = 10~\mu{\mbox A}$ and $C = 3.9 \times 10^{-14}~\mbox{F}$
(the estimate obtained at
the end of Sec.~\ref{sec:static}), eq.~(\ref{T=0}) gives
$S_{inst} \approx 460 (1 - \bmi_b)^{5/4}$. This estimate suggests that, even for this
large
value of $C$, raising the current to within
10\% of the critical will make the rate large enough for quantum phase slips to become 
observable.

\section{Discussion}
\label{sec:conc}
In this paper, we have looked at both classical and quantum 
mechanisms of decay of supercurrent in superconducting nanowires. These mechanisms 
correspond, respectively, to over-barrier activation and tunneling. For the former,
our main conclusion is that the power-3/2 
scaling law (\ref{3/2}) for the activation barrier, often observed experimentally,
is readily reproduced in a discrete model of the wire. That is so even though we
assume
that the values of the phase at the ends remain unchanged during the
activation process (a boundary condition attributed to the presence of bulk 
superconducting leads) and keep $N$, the length of the wire, finite and possibly small 
when sending the current to the critical.

Next, we have found that, in this discrete model, 
the crossover to the quantum regime occurs in a continuous 
manner similar to a supercritical Hopf bifurcation. We have found numerically
the Euclidean solutions (periodic instantons) that describe tunneling for temperatures
ranging from $T$ just below the crossover $T_q$ to $T$ close to zero. We have also observed
that the slowing down of the Euclidean dynamics near the critical current, leads
to the power-5/4 scaling law for the tunneling exponent at $T \ll T_q$.

Physically, the discrete model represents the idea that the spatial size of a phase slip
is determined by the size of a Cooper pair (Pippard's coherence
length in the clean limit or its counterpart, discussed in Sec.~\ref{sec:model}, in the 
dirty limit). Such a phase slip will appear point-like in any local theory that deals
with the order 
parameter alone, for instance, in the standard GL theory. In other words, the 
activation path described here is physically distinct from that mediated by the LA 
saddle point \cite{LA} of the GL model (and it is, then, perhaps not surprising that 
the activation barrier follows a different scaling law).

Finally, let us return to the question asked in the beginning of this paper, namely, 
whether it is possible to represent the results obtained so far
as consequences of the
dynamics of a single phase variable. We note at once that this variable cannot be 
the phase
difference between the ends of the wire, because we assume that the phases at the ends
do not change at all during either activation or tunneling. 
A variable that does look suitable is the jump,
$\vartheta_{core} = \theta_{k+1} - \theta_k$, of the phase across the core of a phase
slip. This variable is ``emergent,'' in the sense that it refers specifically to
solutions describing phase slips: critical droplets and periodic instantons.
In Sec.~\ref{sec:comp},
we have seen that, at large $N$ and low currents, $\vartheta_{core}$ changes
during tunneling
by nearly the full $2\pi$. On the other hand, it changes by only 
a small amount at currents near the critical. These properties 
are consistent with the intuition 
about how the right variable should behave. We expect that a phenomenological theory
of $\vartheta_{core}$, based on a suitable effective potential,
will be able to
reproduce the various scaling laws discussed in the present paper.

The author thanks A. Bezryadin for comments on the manuscript.

\appendix
\section{Activation barrier in the general case}
Here, we generalize the results of Sec.~\ref{sec:static} to the case when the current
in each link is a periodic but non-sinusoidal function of the phase difference:
\be
I_{j,j+1} = F'(\theta_{j+1} - \theta_j) \, .
\ee
We assume that $F'(\vartheta)$ is the derivative of a smooth $2\pi$-periodic function
$F(\vartheta)$ (the potential) 
and is subject only to the following two conditions. (i) $F'(\vartheta)$ is odd
(a consequence of time-reversal invariance). This implies $F'(0) = 0$ and, due to
the periodicity, also $F'(\pi)=0$. (ii) $F'(\vartheta)$ has exactly one maximum (and
no minima) on $(0,\pi)$. The maximum value of $F'$ is the critical current $I_c > 0$. 

Condition (ii) implies that for any
$0\leq I < I_c$, the equation $F'(\vartheta) = I$ has exactly 
two roots on $[0,\pi]$. We will use the following notation: if one of the roots is
$\vartheta_1$, we will write the other as
\be
\vartheta_2 = \vartheta_1^\vee
\ee
and refer to it as the coroot of $\vartheta_1$. As an example, 
for the sinusoidal $F'$, $\vartheta^\vee = \pi - \vartheta$.

In the ground state corresponding to biasing current $I_b$,
all the phase differences $\th_{j+1} -\theta_j$ are equal to $\vgs$, the smaller root of 
$F'(\vgs) = I_b$. For the critical droplet, all of them except one are equal
to $\vgs+\gamma$, and the remaining one to $(\vgs+\gamma)^\vee$. By our choice of the boundary 
conditions, the total phase difference $\th_N - \th_0$
for the droplet must be the same as for the ground state. This leads to the following
equation for $\gamma$:
\be
(N-1) \gamma + (\vgs+\gamma)^\vee = \vgs \, .
\label{eq_gamma_gen}
\ee
In the limit when $N \to \infty$ with $\vgs$ fixed, there is a unique solution:
\be
\gamma = - \frac{1}{N} \left( \vgs^\vee - \vgs \right) + O(1/N^2) \, .
\ee
We will assume that, even if $N$ is not particularly large, it remains
large enough for the solution to exist and be unique.
(For the sinusoidal $F'$, this leads to the condition $N > 2$ mentioned
in the main text.)

The energy of the critical droplet, relative to 
the ground state, is 
\be
E_{sph} - E_{gs} = (N-1) F(\vgs+\gamma) + F[(\vgs+\gamma)^\vee] - N F(\vgs) \, .
\label{Esph_gen}
\ee
Special cases arise when $\gamma$ is small, and we can expand (\ref{Esph_gen}) in it.
Expanding to the second order and using (\ref{eq_gamma_gen}) and the definition of
the coroot, we obtain
\be
E_{sph} - E_{gs} = F(\vgs^\vee) - F(\vgs)  + I_b (\vgs - \vgs^\vee) + O(N F'' \gamma^2) \, .
\label{Esph_lin}
\ee
The second derivative in the last term is taken at $\vgs$.
At $N \gg 1$, $\gamma$ is $O(1/N)$, so in the limit $N\to \infty$ (\ref{Esph_lin})
coincides perfectly with the activation energy of a Josephson
junction whose potential is $U(\phi) = F(\phi) - I_b \phi$.

Next, consider $I_b$ close to the critical current $I_c$ (with $N$ fixed). 
In this case, $\vgs^\vee - \vgs \equiv 2 \eps$ is small, and $\gamma$ is
of order $\eps$. Upon the expansion
\be
F(\vgs^\vee) - F(\vgs) = I_b (\vgs^\vee - \vgs) + O(\eps^3) \, ,
\label{eps3}
\ee
we see that the $O(\eps)$ terms in (\ref{Esph_lin}) cancel.
The last term in (\ref{Esph_lin}) is $O(\eps^3)$, because near the maximum of $F'$
the second derivative $F''$ is small, here of order $\eps$. Since 
$\eps$ scales as $(1 - I_b/I_c)^{1/2}$, we obtain, for this general case, the same
power-3/2 scaling (\ref{3/2}) as found earlier for the sinusoidal $F'$.

\end{document}